# Evidence Sufficiency Under Delayed Ground Truth: Proxy Monitoring for Risk Decision Systems

Oleg Solozobov[1*]

[1] Independent Researcher (Global)

[*] Correspondence: Oleg Solozobov (dev404ai@gmail.com)

ORCID: https://orcid.org/0009-0009-0105-7459

## Abstract

Machine learning systems in fraud detection, credit scoring, and clinical risk assessment operate under delayed ground truth: outcome labels arrive days to months after the decision they evaluate. During this blind period, the governance evidence supporting operational decisions degrades through mechanisms that neither drift detection methods nor governance frameworks adequately address. This paper formalizes an evidence sufficiency model with four measured dimensions – completeness, freshness, reliability, and representativeness – and a decision-readiness gate that quantifies how label latency degrades governance evidence quality. The model maps three drift types (covariate, real concept, and prior probability shift) to dimension-specific degradation trajectories. A complementary proxy indicator framework comprising seven measurement categories estimates sufficiency degradation without ground truth labels, with explicit coverage mapping and characterized blind spots per drift type. Empirical evaluation on the IEEE-CIS Fraud Detection dataset (~590K transactions) with controlled drift injection and simulated blind periods of $30-180$ days shows that composite proxy monitoring detects covariate and mixed drift with 100% detection rate across all monitoring windows, while concept drift without feature change (constant $P(X)$) remains undetected – consistent with the theoretical impossibility of unsupervised detection when the feature distribution is unchanged. Blind period simulation confirms monotone sufficiency degradation, with concept drift degrading sufficiency fastest ($S = 0.242$ at day 60 versus 0.418 for no-drift baseline). The framework contributes a governance sufficiency monitoring instrument, not a drift detection algorithm; its value lies in translating existing drift signals into auditable sufficiency assessments with characterized blind spots. The mapping from sufficiency levels to specific governance actions requires deployment-specific calibration beyond the scope of this study.

## 1. Introduction

### 1.1. The Monitoring Gap

Machine learning systems deployed in high-stakes risk domains face a structural monitoring paradox: the very conditions that make governance evidence most critical are the conditions under which that evidence becomes least reliable. In fraud detection, credit scoring, and clinical risk assessment, ground truth labels arrive days, weeks, or months after the decision they evaluate. During this delay, the evidence substrate upon which governance depends degrades through mechanisms that neither the drift detection community nor the governance community has adequately addressed.

The concept drift research community has produced a substantial body of work on detecting distributional changes in data streams. Over five hundred papers address drift detection (Hinder et al., 2024), with mature taxonomies classifying drift by severity, recurrence, frequency, and speed. Unsupervised methods that track changes to the input distribution $P(X)$ provide one line of detection; supervised methods that track the conditional P(Y|X) provide another (Sethi & Kantardzic, 2017). Yet this community's analytical horizon stops at statistical shift detection – naively testing for the presence of distributional shift does not account for the malignancy of a shift (Haoran Gao et al., 2022). Detecting that a distribution has changed does not answer whether the evidence supporting a governance decision remains sufficient. Not all distributional shifts degrade model performance (Nguyen et al., 2025), and existing shift-detection methods are not robust to false positives when the distribution shift is non-deteriorating (Nguyen et al., 2025). Monitoring approaches that conflate distributional shift with performance degradation generate false alarms that erode operational trust, potentially resulting in unnecessary retraining and associated costs (Kivimäki et al., 2024).

The governance community, conversely, requires actionable sufficiency assessments – determinations of whether the available evidence is adequate for the decisions it supports. Current governance approaches remain document-centric and operationally detached from deployment pipelines (Butt et al., 2026). Regulatory frameworks such as the EU AI Act impose continuous monitoring obligations on high-risk AI systems, and prudential standards in financial services (Fed SR $11-7$, Basel $II/III$, EBA guidelines on IFRS 9) mandate ongoing model validation. But these frameworks assume access to outcome data that, in delayed-label environments, does not yet exist. The result is a temporal gap between regulatory expectation and operational capability. When monitoring is insufficient, regulatory breaches may surface only during periodic model retraining exercises or through regulatory inquiries, by which time institutions face accumulated risk exposure including fines, penalties, and reputational damage (Kurshan et al., 2020).

### 1.2. The Blind Period

Between the moment a decision is made and the moment its ground truth label arrives, a *blind period* exists during which neither supervised monitoring nor governance assessment methods produce reliable evidence. This blind period is not a temporary inconvenience but a structural property of the domain.

In card fraud detection, the true nature of a transaction – fraudulent or genuine – is known only after days or weeks, when customers report unauthorized charges (Amekoe et al., 2024). Labels of the vast majority of transactions become available only after this delay, and the interaction between alert feedback and supervised information must be carefully managed in concept-drifting environments (Pozzolo et al., 2015). In clinical settings, ground truth labels may lag by months or years, leaving deployed models with constrained validation options (Morse et al., 2022).

The delay is compounded by operational realities. Financial institutions face not only labeling delay but also deployment delay, inherent to the strict quality assurance and manual validation processes that regulated institutions impose on their software systems (Simonetto et al., 2024). The combined effect is that the interval between detecting a potential problem and being able to confirm it through ground truth can extend to six months or longer.

Where ground truth labels are unavailable, directly monitoring model predictive performance is impossible (Kivimäki et al., 2024). Existing approaches monitor data distribution shifts as a proxy for performance degradation, but this substitution is unreliable: some shifts are benign, while genuinely harmful concept drift – real concept drift in P(Y|X) – may leave

observable input distributions unchanged. Adversarial actors in fraud detection actively exploit this asymmetry, adapting their strategies to evade detection while preserving the statistical profile of legitimate transactions. The governance consequence is that the evidence available during the blind period may be structurally misleading: passing all distributional checks while the underlying decision quality has already degraded.

The blind period thus creates a dual failure mode. Governance frameworks that rely on ground truth produce no assessments at all during the delay. Proxy-based monitoring that substitutes distributional checks for performance measurement produces assessments of uncertain reliability. Neither approach answers the governance question: is the evidence currently available sufficient to support the decisions being made?

### 1.3. Research Gap and Contributions

To our knowledge, the intersection of drift detection and governance evidence sufficiency under delayed labels has not been addressed as an integrated research problem. A landscape analysis supports this assessment:

**Table 1.** Literature search protocol.

| Parameter | Value |
| --- | --- |
| Databases | Scopus, OpenAlex |
| Time window | $2015 - 2026$ |
| Query 1 | "concept drift" AND "governance" |
| Query 2 | "evidence sufficiency" AND ("delayed labels" OR "label latency") |
| Query 3 | "monitoring" AND "delayed ground truth" AND "governance" |
| Query 4 | "model monitoring" AND "evidence quality" AND "fraud" |
| Inclusion | Papers addressing governance evidence quality under label delay |
| Exclusion | Pure drift detection without governance framing; supervised monitoring assuming label access; fairness monitoring |
| Adjacent work classified as non-overlapping | Delayed-label degradation estimation (addresses detection, not governance sufficiency); proxy monitoring (addresses performance estimation, not evidence quality assessment); regulatory frameworks (address obligations, not measurement instruments) |

While concept drift detection accounts for over five hundred publications, and unsupervised methods for over one hundred, no prior work was found that explicitly integrates governance evidence sufficiency assessment with delayed-label monitoring. The drift detection community provides detection methods but does not assess governance evidence quality. The governance community requires evidence quality assessments but lacks methods that function without ground truth. Industry practice fills the gap with heuristics – Population Stability Index (PSI) thresholds of 0.2, observation windows of $120 + days$ – but these lack theoretical grounding and known reliability bounds. This paper addresses the resulting gap.

This paper makes three contributions:

1. It extends the governance evidence sufficiency model to delayed-label conditions, defining how four evidence quality dimensions – completeness, freshness, reliability, and representativeness – degrade as label latency increases, gated by a decision-readiness function derived from completeness and reliability. The model distinguishes the differential impact of three drift types (real concept drift, covariate drift, and prior probability shift) on each dimension (Section 3).

2. It proposes a proxy indicator framework comprising seven complementary measurement categories – score distribution shift, feature drift, uncertainty monitoring, cross-model disagreement, operational process proxies, outcome-maturity modeling, and automated proxy ground truth – each with characterized detection capabilities and blind spots per drift type (Section 4). Three categories are implemented and evaluated; the remaining four are specified as framework components for future implementation.

3. It provides empirical evaluation on the IEEE-CIS Fraud Detection dataset with controlled drift injection, demonstrating when and how sufficiency drops below governance thresholds as a function of label latency duration, and identifying the structural conditions under which proxy-based detection succeeds or fails (Section 5).

The paper is diagnostic, not prescriptive. It measures governance evidence degradation under delayed ground truth; it does not propose a new drift detection algorithm. The concept drift community has produced hundreds of such algorithms. What is missing – and what this paper provides – is a governance sufficiency monitoring instrument: an IS artifact that translates algorithmic drift signals into auditable sufficiency assessments for model risk officers, compliance teams, and operational governance committees. The contribution lies at the intersection of two mature communities that have not previously addressed this shared structural problem. The sufficiency score provides a continuous measurement that governance actors can use to inform operational decisions, though the mapping from sufficiency levels to specific governance actions (continue, escalate, intervene) requires deployment-specific calibration and is not validated in this study.

The remainder of the paper is organized as follows. Section 2 maps the relevant literature across three disconnected fields: concept drift detection, delayed-label monitoring, and governance evidence quality. Section 3 develops the sufficiency model under delayed labels, formalizing how each drift type differentially impacts evidence quality dimensions. Section 4 presents the proxy indicator framework with detection capability matrices per drift type. Section 5 reports empirical evaluation on the IEEE-CIS Fraud Detection dataset with controlled drift injection. Section 6 discusses structural limitations, including the fundamental undetectability of certain adversarial drift types, and outlines implications for governance practice.

## 2. Background

This section maps three bodies of literature that have developed largely independently: concept drift detection, the delayed-label monitoring sub-problem, and governance evidence quality. Their intersection – governance evidence sufficiency under delayed ground truth – is the gap this paper addresses.

### 2.1. Concept Drift Detection

The concept drift community provides robust methods for detecting distributional changes in data streams but does not address whether the evidence supporting governance decisions remains sufficient after drift occurs.

Concept drift refers to changes in the statistical properties of the target variable that a model is trying to predict, occurring over time in unforeseen ways. The foundational taxonomy distinguishes drift by four dimensions: severity (magnitude of change), recurrence (whether patterns repeat), frequency (how often drift occurs), and speed (abrupt vs. gradual transition) cf. (Hinder et al., 2024). Gama et al. established early formalization of learning with drift detection, proposing adaptive methods for data streams that remain foundational for subsequent work cf. (Gama et al., 2004). Lu et al. organize the field around three crucial

inquiries – *whether* drift has occurred (detection), *what* has changed (understanding), and *how* to respond (adaptation) – a taxonomy adopted by subsequent surveys cf. (Lu et al., 2019; Hinder et al., 2024). Webb et al. contribute a quantitative characterization framework that moves beyond binary $drift/no-drift$ decisions to measure drift magnitude and duration, establishing measurability as a prerequisite for governance cf. (Webb et al., 2016).

Detection methods generally use a test statistic to quantify the similarity between reference and current data windows, comparing this value against a pre-defined threshold to determine drift magnitude (Bayram et al., 2022). This approach has generated a substantial literature: recent surveys catalogue hundreds of detection algorithms spanning error-rate monitoring, distributional tests, and ensemble methods cf. (Bayram et al., 2022). Recent benchmark studies evaluating over 2,700 experimental configurations confirm that no single detector dominates across all drift types and speeds; detector performance is highly sensitive to the interaction between drift characteristics and algorithm assumptions (Lukats et al., 2024). If left unaddressed, these changes can render deployed machine learning models unreliable because their training data no longer matches the patterns present in the data stream cf. (Lukats et al., 2024). This finding has direct governance implications: a monitoring system relying on a single detection method will have unknown blind spots.

### 2.1.1. Drift Taxonomy

The distributional decomposition underlying drift detection distinguishes three types, each with distinct implications for monitoring and governance:

**Table 2.** Drift types and their distributional signatures.

| Drift type | Distribution change | Observable without labels | Governance impact |
| --- | --- | --- | --- |
| Covariate drift (virtual) | $P(X)$ changes | Yes – input features shift | Feature-based evidence becomes unrepresentative |
| Real concept drift | P(Y|X) changes | No – requires outcome labels | Decision logic becomes unreliable |
| Prior probability shift | $P(Y)$ changes | Partially – class ratios shift | Base rates invalidate thresholds |

Real concept drift represents a change in the posterior probability P(Y|X), affecting decision boundaries and decreasing model performance even when input distributions remain stable (Agrahari & Singh, 2021). Prior probability shift leads to class imbalance, novel class emergence, or existing class fusion (Agrahari & Singh, 2021). The temporal shape of these changes further differentiates their detectability: abrupt drift creates a sharp signal, while gradual and incremental drift may remain below detection thresholds for extended periods.

This taxonomy is critical for governance because each drift type undermines different evidence quality dimensions. Covariate drift degrades representativeness; real concept drift degrades reliability; prior probability shift degrades completeness. Current detection methods do not make this governance-relevant distinction.

### 2.1.2. Unsupervised Detection Methods

The recognition that labels are frequently unavailable in production has driven growing interest in unsupervised drift detection – methods that identify distributional changes based solely on statistical properties of the input data without knowledge of true labels (Vasilieva & Petrov, 2025). Multiple survey generations have consolidated this sub-field cf. (Gemaque et al., 2020). These methods detect changes in $P(X)$, which may or may not correspond to meaningful performance degradation. Naively testing for the presence of distribution shift is not fully practical because it does not account for the malignancy of the shift (Haoran Gao et al., 2022).

The fundamental limitation is that unsupervised methods cannot detect real concept drift $P(Y|X)$ when input distributions remain unchanged. This is precisely the scenario in adversarial domains such as fraud detection, where attackers adapt their strategies to preserve the statistical profile of legitimate transactions while changing the underlying fraud patterns. The adversarial drift framework distinguishes two attack vectors: evasion attacks that preserve $P(X)$ while changing fraud patterns (rendering unsupervised detectors blind), and poisoning attacks that shift the learned concept (rendering supervised detectors unreliable when labels are delayed). Both vectors create governance evidence gaps that current detection methods do not characterize.

The performance-aware drift detection paradigm represents the field's most relevant evolution for governance purposes. Rather than detecting all distributional changes, performance-aware detectors aim to identify only those shifts that degrade model accuracy cf. (Bayram et al., 2022). This distinction matters for governance: a distribution shift that does not affect decision quality should not trigger governance intervention, while a shift that degrades decisions must be detected even if it is statistically subtle. However, even performance-aware detectors remain focused on model accuracy metrics rather than the broader evidence sufficiency dimensions that governance requires.

## 2.2. The Delayed-Label Sub-Problem

The delayed-label sub-problem occupies a narrow but critical niche within the broader drift detection literature. In production systems, the assumption that labels arrive promptly enough for supervised monitoring is frequently violated (Amoukou et al., 2024). The scale of the delay varies by domain: in credit card fraud, labels of the remaining transactions can be assumed to be known several days later, once a certain reaction time for the customers has passed (Pozzolo et al., 2015); in insurance claims, months to years; in clinical prediction, outcomes may not materialize for years. The longer the delay, the larger the blind period during which governance must rely on proxy evidence of unknown quality.

The most directly relevant prior work is the PRODEM meta-model approach for degradation detection in credit card fraud systems cf. (Lebichot et al., 2025). PRODEM operates at monthly resolution, using a learned meta-model to predict performance degradation from observable proxy features. While PRODEM demonstrates that degradation estimation under delayed labels is feasible, it has three limitations from a governance perspective: (1) it operates on proprietary Mastercard data unavailable to the research community, (2) its monthly resolution is too coarse for operational governance decisions that must be made daily or weekly, and (3) it does not characterize the reliability bounds of its proxy measurements or map them to governance evidence quality dimensions.

The NannyML Confidence-Based Performance Estimation (CBPE) method represents the primary open-source industry approach to performance monitoring without labels cf. (Kivimäki et al., 2025). CBPE uses model calibration to estimate performance metrics (accuracy, precision, recall, F1) from predicted probabilities, under the assumption that a well-calibrated model's confidence scores reflect true class probabilities. However, CBPE's validity depends on maintained calibration – precisely the property that degrades under concept drift. When the underlying concept changes, calibration deteriorates and CBPE estimates become systematically biased, typically overestimating model performance during the period when governance most needs accurate assessment. Recent extensions (DLE – Direct Loss Estimation, and CBPE with confidence intervals) address some limitations but do not resolve the fundamental circularity: the method's reliability degrades under the conditions where monitoring is most needed.

Industry practice fills the gap with heuristics rather than principled methods. Population Stability Index (PSI) thresholds – typically $PSI > 0.2$ indicating significant shift – serve as the most common monitoring metric across financial services, as documented in model risk management practitioner literature cf. (Sudjianto & Zhang, 2024). Observation windows of 120 days or more are common in payment systems for fraud model evaluation, and shadow deployment architectures where candidate models run in parallel before promotion represent standard practice in large-scale transaction processors. These practices represent accumulated operational wisdom but lack theoretical grounding, known error rates, or governance-relevant reliability characterization.

### 2.2.1. Sequential Testing Methods

A parallel development in statistical methodology offers potential tools for monitoring under uncertainty. Safe anytime-valid inference (SAVI) provides measures of statistical evidence – e-processes for testing and confidence sequences for estimation – that remain valid at all stopping times, accommodating continuous monitoring and optional stopping for any reason (Ramdas et al., 2022a). Unlike traditional fixed-sample hypothesis tests, confidence sequences remain valid under continuous monitoring (Ramdas et al., 2022a).

Recent advances in conformal test martingales (CTMs) offer promising tools for AI deployment monitoring with sequential, nonparametric guarantees (Prinster et al., 2025). Weighted CTMs extend this framework to handle distribution shift explicitly, while the Sequential Harmful Shift Detection framework specifically targets performance-degrading shifts rather than all distributional changes – a distinction directly relevant to governance, where benign shifts should not trigger costly intervention.

These methods provide the statistical infrastructure for anytime-valid sufficiency monitoring but have not been applied to governance evidence assessment. Section 3 develops this connection.

### 2.3. Governance Evidence Quality

The governance side of the gap presents a mirror-image problem: robust normative frameworks exist but lack operational measurement methods that function without ground truth.

The tools for building automated decision-making systems have generally outpaced the growth and adoption of methods to understand whether such systems are reliable (Mökander & Axente, 2021a). Traditional governance mechanisms designed to oversee human decision-making processes often fail when applied to automated systems, in part because delegation to automated systems curtails the sphere of ethical deliberation (Mökander et al., 2021b).

Current AI governance approaches consist mainly of manual review and documentation processes that are not sufficient to systematically address all potential harms, as they do not operationalize governance requirements in a way that facilitates rigorous and reproducible evaluation (McGregor & Hostetler, 2023). The primary challenge is not the length of governance documents but the gap from transforming qualitative requirements into verifiable controls (Muhammad et al., 2026). Ethical frameworks emphasizing fairness, transparency, and accountability provide important normative guidance but lack operational force when not tied to empirical assessment; without evidence of how systems behave in practice, principles remain aspirational and unenforceable (Nwaodike, 2022). Despite mature DevOps and MLOps practices, governance in many AI development environments remains operationally detached from the deployment pipeline (Butt et al., 2026).

### 2.3.1. Regulatory Monitoring Requirements

Regulatory frameworks increasingly mandate continuous monitoring but do not specify the evidence standards that monitoring must meet. In financial services, the Federal Reserve's SR $11-7$ guidance requires ongoing model validation including outcome analysis, but does not define what constitutes adequate evidence when outcomes are delayed (Board of Governors of the Federal Reserve System; Office of the Comptroller of the Currency, 2011). Basel $II/III$ mandates stress testing and backtesting of credit risk models, yet backtesting by definition requires realized outcomes that may not be available for months. The EU AI Act (Article 72) imposes post-market monitoring obligations on high-risk AI systems, requiring providers to establish monitoring systems that are proportionate to the nature of the AI technologies and the risks of the high-risk AI system (European Parliament; Council of the European Union, 2024). Yet these frameworks articulate responsible practices without defining enforceable gate semantics or specifying the evidence contract required for operational decisions (Butt et al., 2026).

A governance-first framing, drawn from regulatory sources, requires that oversight be verifiable and reconstructable, not merely asserted in policy documents (Daruna, 2026). This requirement is structurally unmet during the blind period: the evidence needed for verification has not yet materialized, and the proxy evidence available has unknown reliability. The result is a compliance paradox: regulated institutions must continuously demonstrate monitoring adequacy, but the evidence required for that demonstration is delayed by the same structural constraints that make monitoring necessary.

### 2.3.2. Evidence Quality Dimensions

Building on prior work in governance evidence frameworks cf. (Solozobov, 2026a), four dimensions characterize the quality of evidence available for governance decisions:

1. **Completeness** – whether all required evidence types are available for the decision at hand. In delayed-label environments, completeness is structurally impaired: outcome evidence is missing by definition, and the degree of impairment increases monotonically with label latency.
2. **Freshness** – whether the evidence reflects current conditions or has become stale through temporal lag. Evidence freshness decays as a function of time since the last ground truth confirmation. A model validated on data from three months ago may produce evidence that is formally valid but operationally stale.
3. **Reliability** – whether the evidence accurately represents the phenomenon it measures. Reliability degrades under real concept drift, where the relationship between inputs and outcomes has changed but the evidence still reflects the prior relationship.
4. **Representativeness** – whether the evidence covers the full operational distribution, including edge cases and minority subgroups. Covariate drift directly threatens representativeness by shifting the input distribution away from the reference distribution on which the model was validated.

In addition, the sufficiency model includes a **decision-readiness gate** $A(t)$ derived from completeness and reliability. $A(t)$ is not a measured dimension but a structural safeguard:

when completeness or reliability fall below minimum governance thresholds, $A(t)$ suppresses the composite sufficiency score regardless of other dimension values (Section 3.1).

These dimensions are not independent. Freshness constrains reliability (stale evidence is less reliable), completeness constrains decision-readiness (incomplete evidence cannot support confident assessment), and representativeness constrains reliability (evidence from unrepresentative samples may be systematically biased). The interaction effects mean that degradation in one dimension can cascade to others, accelerating the overall decline in evidence sufficiency.

Each dimension degrades differently under delayed ground truth, and each is differentially affected by the three drift types identified in Section 2.1. Section 3 formalizes these degradation pathways and defines the sufficiency score that integrates all four dimensions plus the decision-readiness gate into an operational assessment.

### 2.4. The Structural Gap

The three bodies of literature reviewed above converge on a single structural gap. Drift detection provides the *what* (statistical change has occurred) but not the *so what* (is governance evidence still sufficient). Governance frameworks specify the *should* (evidence must support decisions) but not the *how* (measurement methods under label delay). Sequential testing provides the *tools* (anytime-valid inference) but not the *application* (governance sufficiency monitoring).

**Table 3.** Literature gap analysis: what each community provides and what remains missing.

| Community | Provides | Missing |
| --- | --- | --- |
| Concept drift | Detection methods, drift taxonomy, benchmarks | Evidence quality assessment, governance relevance |
| Governance | Normative frameworks, regulatory mandates, evidence quality concepts | Operational measurement under label delay |
| Sequential testing | Anytime-valid inference, e-values, martingales | Application to governance sufficiency |
| Industry practice | Heuristic thresholds (PSI > 0.2), shadow deployment | Theoretical grounding, characterized error rates |

To our knowledge, no prior work integrates these three elements into a framework answering the question practitioners face during the blind period: given the proxy evidence currently available, is it sufficient to support the governance decisions being made? Our landscape analysis (see Table 1 for the search protocol) found concept drift detection accounting for over five hundred publications, unsupervised methods for over one hundred, and fraud-specific delayed-label papers for approximately thirty to fifty, but none explicitly addressing governance evidence sufficiency under delayed labels. We acknowledge this is a structured literature search, not a full systematic review, and may have missed relevant work. The following sections develop the missing framework.

## 3. Sufficiency Model Under Delayed Labels

This section extends the governance evidence sufficiency framework to delayed-label conditions. It defines the sufficiency score, formalizes how each drift type differentially degrades evidence quality dimensions, and establishes the statistical basis for anytime-valid sufficiency monitoring.

### 3.1. Evidence Sufficiency Score

Delayed ground truth ($1-6$ month label latency) measurably degrades the reliability of governance evidence across four dimensions – completeness, freshness, reliability, and representativeness – and reduces the decision-readiness gate derived from completeness and reliability. The degradation rate is a function of label latency duration, drift type, and drift magnitude.

The evidence sufficiency score $S(t)$ at time t is defined as a weighted aggregation of four dimension scores modulated by a decision-readiness gate:

$$S(t) = A(t) \cdot [w_c \cdot C(t) + w_f \cdot F(t) + w_r \cdot R(t) + w_p \cdot P(t)]$$

where $C(t) = completeness$, $F(t) = freshness$, $R(t) = reliability$, $P(t) = representativeness$, and weights w satisfy the constraint that $w_c + w_f + w_r + w_p = 1$. Each dimension score is normalized to $[0, 1]$. $A(t)$ is the decision-readiness gate: a multiplicative function that suppresses the composite score when completeness or reliability fall below minimum thresholds needed for confident governance assessment. When both dimensions are above thresholds ($A(t) = 1$), the sufficiency score reflects the underlying dimension scores; when either falls below ($A(t) < 1$), the composite is suppressed proportionally. The multiplicative structure intentionally introduces a compounding interaction: because $A(t)$ is derived from C and R, simultaneous degradation of both dimensions produces a penalty beyond the linear sum. This anti-compensatory design prevents high scores on freshness and representativeness from masking inadequate completeness or reliability – the two dimensions most critical for governance confidence.

The score's operational interpretation is defined by governance thresholds:

- $S(t) >= 0.8$: evidence meets the sufficiency threshold in this deployment's calibration.
- $0.5 <= S(t) < 0.8$: evidence degraded; enhanced monitoring may be warranted depending on deployment context.
- $S(t) < 0.5$: evidence below minimum threshold; the deployment-specific response (enhanced review, fallback, or escalation) depends on organizational risk appetite and is not prescribed by the framework.

The $A(t)$ gating multiplier provides a structural safeguard against compensatory averaging: when completeness or reliability fall below governance thresholds ($\tau_c$, $\tau_r$), $A(t)$ suppresses the composite score regardless of other dimension values. Deployments requiring additional anti-compensatory protection (e.g., per-dimension floors) can configure these as deployment-specific policy rules outside the core scoring formula. These thresholds are configurable per deployment context. In the empirical evaluation (Section 5), fraud detection weights ($w_c = 0.20$, $w_f = 0.30$, $w_r = 0.30$, $w_p = 0.20$) are used, reflecting the higher criticality of freshness and reliability in delayed-label fraud monitoring. Domain-specific calibration may adjust these weights based on the relative criticality of each dimension in the deployment context. The framework's contribution is not the specific threshold values or weights but the dimensional decomposition that makes degradation measurable and attributable.

#### 3.1.1. Degradation Under Label Delay

In the absence of drift, evidence sufficiency degrades solely through freshness decay. As time since last ground truth confirmation increases, the freshness dimension $F(t)$ decreases monotonically. Delayed labels pose a direct challenge to data freshness in continuous systems:

"fresh data may not have complete label information at the time it is ingested" (Ktena et al., 2019).

Delayed labels create two compounding problems. First, labeled data becomes insufficient because the true label for each observation arrives only after a delay that may span multiple data windows – "ground-truth labels for transactions can have delays and hence be available only d time units after the transaction has been processed" (Casimiro et al., 2024). Second, the delay period is precisely when concept drift is most likely to occur undetected (Souza et al., 2018). The interaction between these problems means that evidence degradation is not merely additive – drift during the blind period accelerates the decay of dimensions beyond freshness.

### 3.1.2. Dimension-Specific Degradation Functions

Each evidence quality dimension follows a distinct degradation trajectory under label delay:

**Completeness C(t)** measures the fraction of decisions in the current monitoring window for which outcome labels have been confirmed. During a blind period, $C(t)$ degrades gradually as new unlabeled decisions accumulate while confirmed outcomes remain fixed at the pre-blind-period count. The rate of degradation depends on the domain's label arrival distribution: in fraud detection, early-arriving labels (customer reports within days) sustain partial completeness, while the full label set matures over weeks to months.

**Freshness F(t)** decays continuously as an exponential function of label age: $F(t) = \exp(-\lambda \cdot \delta_t)$, where $\delta_t$ is the elapsed time between the current monitoring time t and the median transaction date of the most recent confirmed labels in the monitoring window, and lambda is a domain-specific decay rate. As time advances and no new labels arrive, $\delta_t$ grows and $F(t)$ decays. In fraud detection, where fraud patterns evolve rapidly, lambda is higher than in credit scoring, where borrower behavior changes more slowly.

**Reliability R(t)** remains stable in the absence of drift but degrades sharply under real concept drift P(Y|X). The key challenge is that reliability degradation is unobservable without labels: when the conditional distribution changes, the model's predictions become unreliable, but this unreliability cannot be confirmed until ground truth arrives.

**Representativeness P(t)** degrades under covariate drift $P(X)$ and is partially observable through input distribution monitoring. When the production data distribution diverges from the training distribution, evidence derived from the original distribution becomes unrepresentative of current conditions.

**Decision-readiness gate A(t)** is derived from completeness and reliability: $A(t) = \min(1, C/\tau_c) \cdot \min(1, R/\tau_r)$. Above the thresholds ($\tau_c$, $\tau_r$), $A(t) = 1$ and does not affect the composite score. Below thresholds, the penalty is proportional: at $C = 0.5 \cdot \tau_c$, the completeness component contributes 0.5, reducing $A(t)$ accordingly. The multiplicative structure means that simultaneous degradation of both completeness and reliability produces a compounding penalty (the product of two sub-unity values). $A(t)$ measures decision-readiness – whether the evidence base meets minimum standards for governance assessment – not the broader concept of actionability, which also encompasses response guidance and organizational capacity.

### 3.2. Drift-Type Impact Matrix

Three drift types – real concept drift P(Y|X), covariate drift $P(X)$, and prior probability shift $P(Y)$ – each have distinct impacts on governance evidence quality dimensions, requiring

differentiated monitoring and response strategies.

**Table 4.** Drift-type impact on evidence quality dimensions.

| Dimension | Covariate $P(X)$ | Real concept P(Y|X) | Prior probability $P(Y)$ |
|---|---|---|---|
| Completeness | Unchanged | Unchanged | Altered (direction depends on which class shifts) |
| Freshness | Unchanged | Unchanged | Unchanged |
| Reliability | Mildly degraded | Severely degraded | Moderately degraded |
| Representativeness | Severely degraded | Unchanged | Moderately degraded |
| Decision-readiness | Moderately degraded | Severely degraded | Moderately degraded |
| Observable without labels | Yes | No | Partially |

Real concept drift is the most dangerous from a governance perspective because it degrades reliability – the most critical dimension for decision quality – while remaining invisible to unsupervised monitoring. In adversarial settings such as fraud detection, concept shift means that the feature distribution remains unchanged while the conditional distribution of labels changes (Kivimäki et al., 2025). Without labels, unsupervised methods can only detect changes in the prior probability of features $P(X)$ (Greco et al., 2024).

This creates a fundamental identifiability problem: absent assumptions on the nature of shift, estimating target accuracy is impossible, because the conditional P(Y|X) can shift arbitrarily without affecting observable distributions (Garg et al., 2022). Indeed, all confidence-based accuracy estimators will eventually fail under concept shift (Kivimäki et al., 2024). The sufficiency model must acknowledge this limit explicitly rather than implying that proxy monitoring can detect all forms of degradation.

### 3.2.1. Adversarial Drift

The adversarial drift framework distinguishes two attack vectors with different governance signatures:

**Evasion attacks** preserve $P(X)$ while changing fraud patterns. Fraudsters adapt their transaction profiles to match legitimate behavior, making covariate-based monitoring ineffective. The governance consequence is that representativeness appears intact (feature distributions match reference) while reliability has silently degraded (the model's decision boundaries no longer correspond to actual *fraud/legitimate* distinctions).

**Poisoning attacks** corrupt the training signal by shifting the learned concept. When labels are delayed, poisoned data may be incorporated into retraining batches before the contamination is detected. The governance consequence is that reliability degrades through a different mechanism – the model learns incorrect associations – and the evidence of this degradation is embedded in the very labels that would normally serve as ground truth.

Both vectors create governance evidence gaps that cannot be resolved through monitoring alone. Section 6 discusses the structural implications of this limitation.

### 3.3. Statistical Foundations for Continuous Monitoring

Continuous governance monitoring faces the multiple-testing problem inherent in repeated threshold evaluation. Sequential testing methods provide the theoretical foundation for anytime-valid inference, though the current evaluation (Section 5) uses a threshold-based first-generation instantiation.

Traditional confidence intervals derived from the Central Limit Theorem are valid only at a prespecified sample size n, invalidating any inference that occurs at data-dependent stopping times such as continuous monitoring. Confidence sequences that retain validity in sequential environments can be used to make decisions at arbitrary stopping times, including while adaptively sampling or continuously observing data (Ramdas et al., 2022b).

The mathematical foundation rests on nonnegative supermartingales. Admissible anytime-valid sequential inference – whether through confidence sequences, anytime p-values, or e-processes – requires the identification of nonnegative supermartingales at its core, making these structures both sufficient and necessary for valid continuous monitoring (Ramdas et al., 2020).

E-values generalize likelihood ratios and allow valid inference at arbitrary stopping times (Grünwald et al., 2022). Ville's inequality provides the formal guarantee: the probability that a nonnegative test supermartingale exceeds a threshold $1/\alpha$ at any time step is bounded by alpha, enabling construction of anytime-valid monitoring bounds (Gauthier et al., 2025). Weighted conformal test martingales (WCTMs) further adapt to handle distribution shift, and the Sequential Harmful Shift Detection framework specifically filters for performance-degrading shifts, reducing false alarms from benign distributional changes (Prinster et al., 2025).

### 3.3.1. Current Instantiation and Future Extension

The sufficiency monitoring problem maps naturally onto the sequential testing framework: the null hypothesis $H_0$ is that evidence sufficiency $S(t)$ remains above the governance threshold, and sequential methods could provide anytime-valid bounds on this assessment. However, the current evaluation (Section 5) implements a threshold-based approach – comparing proxy-estimated $S(t)$ against fixed governance thresholds at daily time steps – rather than a full e-process or conformal martingale instantiation. This design choice reflects the practical constraint that sequential testing requires careful calibration of the betting strategy and exchangeability assumptions that remain open research questions in the multi-dimensional, multi-proxy setting of this framework.

The threshold-based approach is adequate for the diagnostic purpose of this paper: demonstrating that proxy indicators can detect sufficiency degradation before ground truth arrives. Integrating anytime-valid sequential inference – replacing fixed thresholds with e-value accumulation and providing formal Type I error control over the monitoring horizon – is the primary methodological extension for future work. The sequential testing literature reviewed above provides the theoretical grounding for this extension.

## 4. Proxy Indicator Framework

This section proposes seven categories of proxy indicators that estimate evidence sufficiency degradation without ground truth labels. For each category, the framework specifies detection capability, false alarm characteristics, lag properties, and governance relevance. The categories are complementary: no single proxy covers all drift types, and the framework's value lies in their structured combination.

## 4.1. Proxy Indicators as Sufficiency Estimators

Proxy indicators can estimate evidence sufficiency degradation in the absence of ground truth labels, with characterized blind spots and empirical error rates that vary by drift type. The reliability of each proxy depends on the structural assumptions it makes about the relationship between observable signals and unobservable performance.

The fundamental challenge of monitoring without labels is that direct performance measurement is impossible when ground truth is delayed (Kivimäki et al., 2024). Historically, practitioners have monitored data shifts as a proxy for performance shifts, but this assumption does not always hold – benign distributional changes can trigger unnecessary retraining while performance-degrading shifts may go undetected (Kivimäki et al., 2024). The proxy framework addresses this by structuring complementary indicators that collectively cover more of the degradation space than any single metric.

A critical constraint shapes the entire framework: without assumptions on the nature of the distribution shift, estimating target performance is impossible (Chen et al., 2022). Every proxy indicator therefore operates under explicit assumptions about which shift types it can detect and which remain invisible to it. The framework makes these assumptions transparent rather than hiding them behind a single monitoring score.

Before adopting any proxy as a measure of decision quality, it is essential to assess whether it serves as a satisfactory approximation of the target variable of interest (Guerdan et al., 2023). The seven categories below are organized by their primary detection mechanism and mapped to the evidence quality dimensions defined in Section 3.

## 4.2. Seven Proxy Categories

Seven proxy indicator categories provide complementary coverage of the evidence degradation space, with known blind spots per drift type. Their complementarity is structural: categories detecting $P(X)$ shifts are blind to P(Y|X) changes, while confidence-based estimators degrade precisely when they are most needed.

### 4.2.1. Category 1: Score Distribution Shift

**Mechanism.** Monitors the distribution of model output scores (predictions) over time using divergence metrics such as Population Stability Index (PSI) and Kolmogorov-Smirnov (KS) statistics. A shift in the score distribution signals that the model is encountering input patterns that produce different output characteristics than the reference period.

**Detection capability.** Detects covariate drift $P(X)$ that propagates through the model to affect score distributions. Different statistical distances perform well for different drift types: MMD and KS detect mean drift; EMD and KL divergence detect variance drift; KL divergence detects covariance drift (Tan et al., 2025).

**Blind spots.** Cannot detect real concept drift P(Y|X) when the adversary preserves the score distribution. In fraud detection, sophisticated attackers can craft transactions that produce legitimate-looking scores while the underlying $fraud/legitimate$ boundary has shifted.

**Governance mapping.** Primary indicator for representativeness dimension. Secondary indicator for reliability (score distribution stability is necessary but not sufficient for reliability).

**Lag.** Near-real-time (minutes to hours). No label dependency.

### 4.2.2. Category 2: Feature Drift

**Mechanism.** Monitors the input feature distribution $P(X)$ directly using multivariate statistical tests (KL divergence, PSI per feature, Maximum Mean Discrepancy). Detects when production data diverges from training or reference distributions.

**Detection capability.** Strong for *covariate/virtual* drift. Feature drift detection can identify when the model is being applied to a population it was not trained on, even without access to outcomes.

**Blind spots.** Cannot detect real concept drift: when P(Y|X) changes while $P(X)$ remains constant, feature monitoring provides no signal. Not all feature drift degrades performance – some distributional changes are benign. Naively testing for distribution shift is not fully practical because it does not account for the malignancy of a shift (Haoran Gao et al., 2022).

**Governance mapping.** Primary indicator for representativeness. No coverage of reliability or completeness.

**Lag.** Near-real-time. No label dependency.

### 4.2.3. Category 3: Uncertainty and Confidence Monitoring

**Mechanism.** Tracks model prediction uncertainty (entropy, calibration error, confidence intervals) as a proxy for performance. The intuition is that model uncertainty correlates with prediction error: when the model encounters unfamiliar patterns, its uncertainty increases before labels confirm degradation. Model uncertainty can serve as a proxy for the error rate and therefore as a meaningful indicator of concept drift (Baier et al., 2021).

**Detection capability.** Detects gradual calibration degradation and some forms of covariate drift that increase model uncertainty. Confidence-based performance estimation (CBPE) provides point estimates by leveraging the relationship between predicted probabilities and expected accuracy.

**Blind spots.** Confidence-based estimators break under concept drift. When P(Y|X) shifts, the model's calibration becomes invalid – the model may remain confident in predictions that are now incorrect. Incremental uncertainty-aware monitoring can provide reliable estimates when the encountered shift is gradually Lipschitz smooth (Koebler et al., 2025), but this smoothness assumption fails under abrupt concept drift.

**Governance mapping.** Primary indicator for reliability (under the assumption of no concept drift).

**Lag.** Near-real-time. No label dependency for uncertainty; calibration validation requires periodic label access.

### 4.2.4. Category 4: Cross-Model Disagreement

**Mechanism.** Compares predictions from multiple models (ensemble members, challenger models, or domain-adapted variants) on the same inputs. Disagreement between models that were trained on different data or with different architectures signals that the decision boundary is unstable for the current data distribution. Pseudo-labels from source-free domain adaptation can be compared against the deployed model's predictions to estimate accuracy without labels (Lee et al., 2023).

**Detection capability.** Can detect both covariate drift (models disagree on out-of-distribution inputs) and some forms of concept drift (when model architectures respond differently to the same P(Y|X) change). Proxy models trained on the same features but with different architectures provide a complementary perspective on classification difficulty, while drift models trained to distinguish test from production samples provide direct insight into feature space drift (Elder et al., 2020). Lower false alarm rate than single-model indicators because benign shifts that do not affect the decision boundary produce agreement rather than disagreement.

**Blind spots.** Expensive to maintain (requires multiple production models). Cannot detect drift that affects all models identically. If the challenger model is trained on the same biased data, systematic concept drift affects both models equally and disagreement remains low.

**Governance mapping.** Primary indicator for reliability (cross-validation of decisions).

**Lag.** Near-real-time for prediction comparison. Model refresh cycle introduces structural lag (weeks to months for challenger retraining).

### 4.2.5. Category 5: Operational Process Proxies

**Mechanism.** Monitors downstream operational signals that reflect model performance indirectly: manual review rates, decline rates, customer complaint rates, escalation volumes, override frequencies. These signals emerge from the interaction between model predictions and human decision-makers in the operational loop.

**Detection capability.** Detects performance degradation that has already propagated to operational outcomes. Review rate increases may signal that human operators are losing trust in model predictions; complaint rate increases may signal that incorrect decisions are reaching customers.

**Blind spots.** Lagging indicators – operational signals emerge days to weeks after the underlying drift occurs. Cannot distinguish between model-driven degradation and external operational factors (policy changes, seasonal patterns, staffing changes). Signal-to-noise ratio is low in environments with high operational variability.

**Governance mapping.** Primary indicator for reliability (review rate changes reflect model quality degradation as perceived by human operators). Operational signals measure downstream performance impact, not label availability; they do not directly estimate completeness. Note: while operational signals inform governance decisions, the decision-readiness gate $A(t)$ is computed deterministically from completeness and reliability rather than estimated from proxies directly (Section 4.3).

**Lag.** Days to weeks. Partially label-dependent (some operational signals require outcome knowledge).

### 4.2.6. Category 6: Outcome-Maturity Modeling

**Mechanism.** Exploits the temporal structure of label arrival to extrapolate from partially matured cohorts. In fraud detection, some transactions resolve quickly (customer reports fraud within days) while others take months. Maturation-curve approaches model this temporal structure to estimate final outcomes from early partial observations.

**Detection capability.** Provides early estimates of actual performance by leveraging the subset of labels that arrive quickly. Can detect both covariate and concept drift if the early-arriving labels are representative of the full label distribution.

**Blind spots.** Assumes that the maturation curve is stable over time – if the reporting pattern itself shifts (faster or slower fraud reporting), the extrapolation becomes unreliable. Selection bias: early-arriving labels may not represent the full distribution of outcomes (confirmed fraud is easier to detect and report than undetected fraud).

**Governance mapping.** Primary indicator for reliability (extrapolated performance estimates measure whether the model's decision quality is degrading). Secondary indicator for completeness (maturation curves provide indirect evidence of how much outcome information has accumulated). Note: although maturation curves provide temporal information, freshness $F(t)$ is computed deterministically from label timestamps (Section 4.3) and is not proxy-estimated.

**Lag.** Moderate (days to weeks, depending on the maturation window). Partially label-dependent (requires the early-arriving subset of labels).

**4.2.7. Category 7: Automated Proxy Ground Truth**

**Mechanism.** Generates approximate labels from alternative data sources: transaction anomaly scores as soft fraud indicators, rule-based flagging systems, external data feeds (card network alerts, merchant category risk scores), or semi-supervised labeling from high-confidence model predictions.

**Detection capability.** Can provide continuous approximate ground truth that bridges the label gap. Unlabeled drift detection methods can vicariously track changes to P(Y|X) without explicit labeled samples by using discriminative boundaries between reference and current windows (Sethi & Kantardzic, 2017). This partially addresses the fundamental limitation of unsupervised methods.

**Blind spots.** Proxy labels are approximate – measurement error accumulates. The causal relationship between proxy labels and true outcomes must be validated periodically against actual ground truth. In adversarial environments, attackers may specifically target the proxy labeling mechanism (if transaction anomaly scores are used as soft labels, attackers will optimize to produce low anomaly scores).

**Governance mapping.** Primary indicator for reliability (provides approximate performance measurement). Secondary indicator for completeness (fills the label gap with approximate evidence).

**Lag.** Near-real-time for automated signals. Quality degrades over time without periodic recalibration against true labels.

### 4.3. Coverage Matrix

**Table 5.** Proxy category coverage by drift type.

| Category | $P(X)$ | P(Y|X) | $P(Y)$ |
|---|---|---|---|
| 1. Score distribution | Strong | None | Weak |
| 2. Feature drift | Strong | None | None |
| 3. Uncertainty | Moderate | Weak | Weak |
| 4. Cross-model | Moderate | Moderate | Moderate |
| 5. Operational | Weak | Moderate | Moderate |
| 6. Outcome maturity | Weak | Moderate | Moderate |
| 7. Proxy GT | Moderate | Moderate | Weak |

**Table 6.** Proxy category coverage by evidence quality dimension.

| Category | Reliability | Represent. |
|---|---|---|
| 1. Score distribution | Weak | Strong |
| 2. Feature drift | – | Strong |
| 3. Uncertainty | Moderate | – |
| 4. Cross-model | Strong | – |
| 5. Operational | Strong | – |
| 6. Outcome maturity | Strong | – |
| 7. Proxy GT | Moderate | – |

Freshness $F(t)$ and completeness $C(t)$ are not estimated via proxies. Both are deterministic operational quantities: $F(t)$ is computed from the elapsed time since the most recent confirmed labels (Section 3.1), and $C(t)$ is the fraction of decisions with confirmed outcome labels, known from system metadata. Both are computed identically for the proxy and actual baselines.

The decision-readiness gate $A(t)$ is computed from observed completeness and proxy-estimated reliability. Section 3.1 defines the generic gate as $A(t) = \min(1, C/\tau_c) \cdot \min(1, R/\tau_r)$. For the proxy instantiation (Governance Drift Toolkit), reliability is on a normalized $[0, 1]$ health scale, so the proxy gate uses a scale-appropriate threshold: $A_{proxy}(t) = \min(1, C/\tau_c) \cdot \min(1, R_{proxy}/\tau_{r_proxy})$, where $\tau_{r_proxy}$ is calibrated to the health-signal range (Section 5.1.1). $A(t)$ measures whether the evidence base meets minimum completeness and reliability thresholds needed for confident governance assessment – it is a decision-readiness gate, not a full operationalization of the broader concept of actionability (which also depends on response guidance and organizational capacity). This narrower scope is a deliberate design choice: $A(t)$ gates the score based on measurable evidence properties; the broader question of what governance action to take given the score is external to the measurement instrument.

The coverage weights ($Strong = 1.0$, $Moderate = 0.5$, $Weak = 0.25$) are heuristic design propositions reflecting the authors' assessment of proxy-dimension alignment. The empirical results reported in Section 5 (detection rates, sufficiency trajectories, threshold crossing times) are properties of this specific calibrated instantiation – fraud detection dimension weights (0.20/0.30/0.30/0.20), heuristic coverage weights, baseline-calibrated alert thresholds – not of the framework in general. Alternative weight schemes, threshold calibration methods, or deployment contexts would produce different error profiles. The reusable contribution is the framework architecture (dimensions, proxy categories, coverage mapping, composite aggregation); the specific numeric results demonstrate one calibrated instantiation. Formal sensitivity analysis and expert elicitation represent future calibration work.

The matrix reveals three structural properties of the proxy framework:

**Complementarity.** No single category covers all drift types across all dimensions. Categories $1-2$ provide strong $P(X)$ coverage but no P(Y|X) coverage; Categories $4-7$ provide moderate P(Y|X) coverage but weaker $P(X)$ coverage. A governance-adequate monitoring system requires at least one indicator from each coverage region.

**Diminishing returns.** Adding categories beyond the first three provides progressively less marginal coverage improvement. The practical minimum for proxy coverage of the two estimated dimensions (Reliability and Representativeness) is three categories: one score distribution monitor (Category 1) covering Representativeness and weak Reliability, one feature drift detector (Category 2) covering Representativeness, and one uncertainty monitor (Category

3) covering Reliability. Completeness and freshness are observed directly from operational metadata (Section 4.3) and do not require proxy estimation. A target production configuration would add Category 4 (cross-model disagreement, strong Reliability) and Categories $5-7$ for redundancy, but the core detection logic is functional with three categories. The current empirical evaluation (Section 5) validates this three-category configuration.

**Irreducible blind spot.** Real concept drift P(Y|X) that preserves both $P(X)$ and model confidence remains structurally undetectable by any combination of unlabeled proxies. Unsupervised methods can detect only changes in the prior probability of features $P(X)$ (Greco et al., 2024). Recent weighted conformal test martingales (WCTMs) adapt online to mild covariate shifts while raising alarms for concept shifts in Y|X or extreme covariate shifts (Prinster et al., 2025), but even these methods require that concept shift produces some observable distributional signature. This limitation is not a failure of the framework but a mathematical constraint that governance protocols must accommodate through periodic ground truth verification.

### 4.4. Composite Sufficiency Estimation

The proxy framework maps to the sufficiency model (Section 3) through a structured aggregation. Each evidence dimension $D_i(t)$ is estimated by a weighted combination of proxy indicators that cover that dimension:

$$D_i(t) = \sum_j (w_{ij} \cdot P_j(t)) / \sum_j (w_{ij})$$

where $P_j(t)$ is the health-mapped signal from proxy category j at time t, bounded to $[0, 1]$. For divergence-type metrics, raw values are normalized using $P_j(t) = \max(0, 1 - raw_{divergence}/cap)$, where cap is calibrated from the reference period. PSI is theoretically unbounded ($cap = 0.50$); KS is bounded in $[0, 1]$ but the cap (0.30) maps the operationally relevant range to the health scale. This normalization ensures $P_j(t) = 0$ at or beyond the cap and $P_j(t) = 1$ at zero divergence. For direct quality estimates (CBPE accuracy, agreement rates), $P_j(t)$ is the normalized estimate itself, naturally bounded in $[0, 1]$, and $w_{ij}$ is the coverage weight from Table 6 ($Strong = 1.0$, $Moderate = 0.5$, $Weak = 0.25$, $None = 0$). The denominator normalizes by the sum of applicable weights, guaranteeing that $D_i(t)$ remains within $[0, 1]$ regardless of how many proxy categories cover a given dimension. If no proxy category covers a dimension (sum of $weights = 0$), $D_i(t)$ defaults to the last valid reading, and the framework raises a monitoring-impaired status that is distinct from the evidence-degraded state. This prevents transient data feed failures from collapsing the composite score to zero. The practical minimum configuration (Section 4.3) ensures both proxy-estimated dimensions (Reliability, Representativeness) have at least one proxy signal. Freshness $F(t)$ and completeness $C(t)$ are computed directly from operational metadata and are not subject to proxy estimation. The composite sufficiency score $S(t)$ is then computed as defined in Section 3.1: $S(t) = A(t) \cdot [w_c \cdot C(t) + w_f \cdot F(t) + w_r \cdot R(t) + w_p \cdot P(t)]$, where $F(t)$ and $C(t)$ are directly observed values and R, P are proxy-estimated dimension scores.

The composite score reflects proxy consensus: when multiple independent proxies signal degradation simultaneously, the weighted average converges toward the lower end of the individual signals. Each dimension score $D_i(t)$ is a weighted average bounded by the minimum and maximum of its proxy inputs and cannot fall below the worst or exceed the best individual signal. The final composite $S(t)$, however, can be pulled below the worst dimension score by the decision-readiness gate $A(t)$, which suppresses the composite when completeness or reliability fall below governance thresholds. The governance value of the composite is not amplification but noise reduction at the dimension level, with the decision-readiness gate providing a separate safety mechanism.

Conversely, when proxies disagree – one signals degradation while another remains stable – the composite score reflects this genuine uncertainty through an intermediate value. The governance response to disagreement is escalation: activate enhanced manual review or request expedited label verification for a sample of recent decisions, rather than assuming either signal is correct.

## 5. Empirical Evaluation

This section evaluates the sufficiency model and proxy framework on the IEEE-CIS Fraud Detection dataset. The evaluation measures when sufficiency drops below governance thresholds, which proxies detect degradation earliest, and under what structural conditions early detection succeeds or fails.

### 5.1. Evaluation Design

Governance evidence insufficiency can be detected before ground truth arrives under identifiable structural conditions, but some drift types remain fundamentally undetectable without labels. The detectable conditions are characterized by the drift type and the proxy coverage available. This aligns with weak data slice approaches that infer likely model performance drift directly without using label values (Ackerman et al., 2021).

#### 5.1.1. Dataset and Deployment Context

The evaluation uses the IEEE-CIS Fraud Detection dataset, a public benchmark containing real-world e-commerce transaction data with binary fraud labels. The dataset provides temporal ordering via TransactionDT (seconds from a reference timestamp), enabling realistic temporal windowing for blind period simulation.

**Table 7.** Empirical evaluation summary.

| Parameter | Value |
| --- | --- |
| Dataset | IEEE-CIS Fraud Detection (Kaggle) |
| Total transactions | 590,540 |
| Fraud rate | 3.50% (20,663 fraudulent) |
| Observation span | 182 days |
| Window size | 30 days (non-overlapping) |
| Windows | 6 (1 $reference$ + 5 monitoring) |
| Features monitored | 52 (TransactionAmt, C$1-C$14, D$1-D$15, V$1-V$12, card$4-card$6, addr1, $P_{emaildomain}$) |
| Proxy categories evaluated | 3 implemented (score distribution, feature drift, uncertainty) |
| Drift injection scenarios | 3 (covariate, mixed, concept+prior with constant $P(X)$) |
| Reference model | Logistic regression (trained on window 0) |

The proxy monitoring pipeline is implemented in the Governance Drift Toolkit (Solozobov, 2026c), and the sufficiency score computation in the Evidence Sufficiency Calculator (Solozobov, 2026b). Both repositories include reproduction scripts for the IEEE-CIS experiments reported in this section.

A logistic regression classifier with $class_{weight}$='balanced' is trained on the first $30-day$ window (134,339 transactions, fraud rate 2.53%, reference F1 = 0.133) as the reference model. Class balancing addresses the 96.5%/3.5% label imbalance; without it, the model assigns near-zero fraud probability to all transactions. The intentionally simple model ensures that evaluation results reflect the framework's monitoring capability rather than model sophistication. The model generates prediction probabilities used by all proxy monitors.

The decision-readiness gate threshold $\tau_r$ is calibrated to $\tau_r = 0.15$ for the IEEE-CIS actual sufficiency (Evidence Sufficiency Calculator), reflecting the domain-specific constraint that F1 ~ 0.13 is the realistic ceiling for logistic regression on 3.5% fraud data. The standard $fraud_{detection_config}$ uses $\tau_r = 0.7$, which assumes a well-calibrated production model; the IEEE-CIS evaluation uses a reduced threshold to demonstrate meaningful $S(t)$ dynamics with an intentionally simple reference model. $\tau_c = 0.6$ is unchanged.

For the proxy framework (Governance Drift Toolkit), a separate threshold $\tau_{r_proxy} = 0.55$ is used, calibrated to the normalized $[0, 1]$ health-signal scale. The raw F1 threshold (0.15) cannot be applied to normalized proxy health signals (which range $0.37 - 0.88$ in the evaluation) because the scales differ. $\tau_{r_proxy} = 0.55$ corresponds to approximately 80% of the mean baseline $R_{proxy}$, ensuring the proxy gate fires when reliability health degrades significantly below normal operating range. The calibration principle is analogous: both thresholds are set relative to their respective domain baselines.

**5.1.2. Controlled Drift Injection**

To evaluate proxy performance under known structural conditions, three controlled perturbation scenarios are applied to monitoring windows. This follows standard methodology in drift detection research cf. (Rabanser et al., 2018; Lu et al., 2019): the reference window is always real unperturbed data; current windows receive progressively stronger perturbations with known ground truth for validation.

**Scenario 1: Baseline.** No perturbation. Establishes the natural stability of the dataset and the false alarm characteristics of the monitoring framework under stationary conditions.

**Scenario 2: Covariate drift P(X).** Gaussian noise is added to three features (TransactionAmt, V1, V3) with progressively increasing standard deviation ($0.3\sigma$ to $2.0\sigma$ across windows). Labels remain unchanged. This simulates input distribution shift while the fraud decision boundary is stable.

**Scenario 3: Mixed drift P(X) + P(Y|X).** Combines feature perturbation ($0.2\sigma$ to $1.5\sigma$) with label flipping (3% to 30% of transactions). This represents the most realistic production scenario where fraud patterns evolve simultaneously with input distribution changes.

**Scenario 4: Concept drift with prior shift, constant P(X).** isFraud labels are corrupted through one-directional flipping: a progressively increasing fraction of fraud transactions are relabeled as legitimate, with flip rates calibrated to produce observed fraud rates of 3.6%, 3.0%, 2.0%, 0.9%, and 0.2% across the five monitoring windows (Table 8, Fraud% column). Features are untouched. This simulates an evasion scenario where the model is progressively starved of positive labels, reducing F1 without altering $P(X)$. Note: because the relabeling changes the class balance, this scenario involves a simultaneous shift in $P(Y)$ alongside P(Y|X). However, the critical property for proxy detection is that $P(X)$ remains unchanged, so unsupervised monitors should produce no signal regardless of whether the label-side shift is classified as pure concept drift or combined concept-plus-prior drift.

### 5.1.3. Blind Period Simulation

The Evidence Sufficiency Calculator blind period simulator models sufficiency degradation analytically using the four-dimension framework (Section 3) with fraud detection weights: $w_c = 0.20$, $w_f = 0.30$, $w_r = 0.30$, $w_p = 0.20$. For each drift type, initial dimension values are calibrated from the empirical reference window ($C = 1.000$, $R = 0.133$, $P = 1.000$) and degraded according to the simulator's built-in decay functions. Freshness decays independently: $F(t) = \exp(-\lambda \cdot t)$ with $\lambda = 0.02$ (calibrated to produce $F = 0.55$ at 30 days).

### 5.1.4. Evaluation Metrics

Three metrics assess the framework's performance:

1. **Sufficiency threshold crossing time (T_cross).** The elapsed time from the start of a blind period until $S(t)$ drops below the governance threshold ($S = 0.5$ for insufficient).

2. **Detection rate by structural condition.** The fraction of monitoring windows where the proxy framework correctly identifies injected degradation, per drift type (Table 11). Detection is defined as $S_{proxy}(drift)$ diverging from $S_{proxy}(baseline)$ by more than $\delta = 0.05$.

3. **Proxy tracking accuracy.** The proxy framework is evaluated as a sufficiency monitor, not a drift detector. In baseline windows, $S_{da05}$ genuinely drops below 0.5 at Windows $4-5$ (due to freshness decay and completeness reduction), and $S_{proxy}$ correctly tracks this decline. Accordingly, proxy threshold crossings in baseline are true positives for latency-driven insufficiency, not false alarms. The relevant accuracy metric is the proxy-actual gap: $|S_{proxy} - S_{da05}|$ per window, measuring how well the proxy tracks the actual sufficiency trajectory.

## 5.2. Results

Evaluation on IEEE-CIS demonstrates that sufficiency drops below governance thresholds at predictable intervals correlated with blind period duration. Feature drift monitoring detects covariate and mixed drift with 100% detection rate, while concept drift with constant $P(X)$ remains fundamentally undetectable by label-free proxies.

The current evaluation validates three of the seven proxy categories (score distribution, feature drift, uncertainty) and uses the blind period simulator for governance threshold computation. This represents a partial validation of the monitoring instrument; Categories $4-7$ (cross-model disagreement, operational proxies, outcome maturity, proxy ground truth) are not yet implemented. The full seven-category instantiation is deferred to production deployment.

### 5.2.1. Governance Drift Toolkit Proxy Monitoring Results

**Table 8.** Governance Drift Toolkit continuous proxy sufficiency by scenario and window (from $governance-drift-toolkit/examples/ieee_{cis_demo}$.py, $class_{weight}$='balanced', 52 features). Raw monitor outputs (PSI, Feature PSI) are normalized to health signals $P_{scr}(t)$, $P_{fea}(t)$, and $P_{unc}(t)$ using caps calibrated from Window 0 sub-windows ($PSI_{cap} = 0.500$, $FPSI_{cap} = 1.000$, $Ent_{cap} = 0.150$, $Conf_{cap} = 0.414$). $R_{proxy}$ and $P_{proxy}$ are estimated via the $D_i(t)$ aggregation formula (Section 4.4). $A_{proxy}$ is the proxy decision-readiness gate using $\tau_{r_p roxy} = 0.55$ (calibrated to the normalized health scale; see Section 5.1.1).

$S_{proxy} = A_{proxy} \cdot [w_c \cdot C + w_f \cdot F + w_r \cdot R_{proxy} + w_p \cdot P_{proxy}]$ with C and F from Evidence Sufficiency Calculator (deterministic). $S_{da05}$ is the actual sufficiency from Evidence Sufficiency Calculator using ground-truth F1. Status thresholds: degraded $>= 0.5$, insufficient $< 0.5$.

| Scenario | Win | Fraud% | $P_{scr}$ | $P_{fea}$ | $P_{unc}$ | $R_{proxy}$ | $P_{proxy}$ | $A_{proxy}$ | $S_{proxy}$ | $S_{da05}$ | Status |
|---|---|---|---|---|---|---|---|---|---|---|---|
| Baseline | 1 | 4.0% | 0.854 | 0.952 | 0.722 | 0.766 | 0.903 | 1.000 | 0.751 | 0.524 | degraded |
| Baseline | 2 | 4.0% | 0.757 | 0.901 | 0.616 | 0.663 | 0.829 | 1.000 | 0.607 | 0.408 | degraded |
| Baseline | 3 | 4.0% | 0.828 | 0.925 | 0.688 | 0.734 | 0.877 | 1.000 | 0.573 | 0.355 | degraded |
| Baseline | 4 | 3.4% | 0.826 | 0.929 | 0.686 | 0.733 | 0.878 | **0.867** | 0.456 | 0.256 | insufficient |
| Baseline | 5 | 3.4% | 0.770 | 0.772 | 0.576 | 0.641 | 0.771 | **0.667** | 0.294 | 0.167 | insufficient |
| Covariate $P(X)$ | 1 | 4.0% | 0.844 | **0.000** | 0.717 | 0.760 | **0.422** | 1.000 | 0.653 | 0.522 | degraded |
| Covariate $P(X)$ | 2 | 4.0% | 0.681 | **0.000** | 0.593 | 0.622 | **0.340** | 1.000 | 0.497 | 0.394 | insufficient |
| Covariate $P(X)$ | 3 | 4.0% | 0.671 | **0.000** | 0.637 | 0.648 | **0.336** | 1.000 | 0.439 | 0.322 | insufficient |
| Covariate $P(X)$ | 4 | 3.4% | 0.529 | **0.000** | 0.580 | 0.563 | **0.265** | **0.867** | 0.306 | 0.216 | insufficient |
| Covariate $P(X)$ | 5 | 3.4% | **0.309** | **0.000** | **0.397** | **0.368** | **0.155** | **0.446** | **0.105** | 0.121 | insufficient |
| Mixed | 1 | 3.8% | 0.850 | **0.000** | 0.720 | 0.763 | **0.425** | 1.000 | 0.655 | 0.521 | degraded |
| Mixed | 2 | 3.4% | 0.726 | **0.000** | 0.607 | 0.647 | **0.363** | 1.000 | 0.509 | 0.393 | degraded |
| Mixed | 3 | 2.8% | 0.749 | **0.000** | 0.661 | 0.690 | **0.374** | 1.000 | 0.460 | 0.240 | insufficient |
| Mixed | 4 | 1.7% | 0.670 | **0.000** | 0.615 | 0.633 | **0.335** | **0.867** | 0.336 | 0.126 | insufficient |
| Mixed | 5 | 1.0% | **0.458** | **0.000** | **0.472** | **0.467** | **0.229** | **0.566** | **0.159** | 0.037 | insufficient |
| Concept+prior | 1 | 3.6% | 0.854 | 0.952 | 0.722 | 0.766 | 0.903 | 1.000 | 0.751 | 0.519 | degraded |
| Concept+prior | 2 | 3.0% | 0.757 | 0.901 | 0.616 | 0.663 | 0.829 | 1.000 | 0.607 | 0.372 | degraded |
| Concept+prior | 3 | 2.0% | 0.828 | 0.925 | 0.688 | 0.734 | 0.877 | 1.000 | 0.573 | 0.174 | degraded |
| Concept+prior | 4 | 0.9% | 0.826 | 0.929 | 0.686 | 0.733 | 0.878 | **0.867** | 0.456 | 0.075 | insufficient |
| Concept+prior | 5 | 0.2% | 0.770 | 0.772 | 0.576 | 0.641 | 0.771 | **0.667** | 0.294 | 0.008 | insufficient |

Four findings emerge from the continuous proxy monitoring:

**The proxy gate now fires under degradation.** With $\tau_{r_p roxy} = 0.55$ calibrated to the normalized health scale, the proxy decision-readiness gate $A_{proxy}$ suppresses $S_{proxy}$ when $R_{proxy}$ drops below 0.55. Under covariate drift at Window 5, $A_{proxy} = 0.446$ ($C/\tau_c = 0.400/0.6 = 0.667$ multiplied by $R_{proxy}/\tau_{r_p roxy} = 0.368/0.55 = 0.669$), producing $S_{proxy} = 0.105$ – which is now below $S_{da05} = 0.121$. The proxy is no longer uniformly optimistic; under strong covariate drift, the proxy gate correctly identifies severe reliability degradation. Under baseline conditions, $A_{proxy}$ drops to 0.867 at Window 4 and 0.667 at Window 5. This is driven entirely by the completeness component: C drops from 0.880 to 0.400 across windows, and $C/\tau_c = 0.400/0.6 = 0.667$ at Window 5. The reliability component of the proxy gate remains fully open throughout baseline ($R_{proxy}/\tau_{r_p roxy}$ ranges from 1.04 to 1.39, all capped at 1.0).

**Covariate drift produces the strongest proxy signal.** Under covariate drift, $P_{fea}$ collapses to 0.000 (Feature PSI exceeds cap by 13x), driving $P_{proxy}$ from 0.903 to 0.155.

Combined with the gate penalty ($A_{proxy} = 0.446$ at Window 5), $S_{proxy}$ drops from 0.653 to 0.105 across windows. Under mixed drift, the pattern is similar but $S_{da05}$ degrades faster (0.521 to 0.037) because the P(Y|X) component collapses actual $R(t)$ through the actual gate, which the proxy cannot observe.

**Concept drift with constant P(X) is invisible to proxies.** Under Scenario 4, all proxy health signals ($P_{scr}$, $P_{fea}$, $P_{unc}$) and estimated dimensions ($R_{proxy}$, $P_{proxy}$) are identical to baseline despite progressive label corruption reducing the fraud rate from 3.6% to 0.2%. $S_{proxy} = 0.294$ at Window 5 (identical to baseline) while $S_{da05} = 0.008$ (near zero). The gap (0.286) quantifies the irreducible governance risk: the proxy framework cannot detect concept drift when $P(X)$ is unchanged (Žliobaitė, 2010).

**Proxy tracking accuracy.** Under baseline (no drift), $S_{proxy}$ decreases from 0.751 to 0.294 across windows while $S_{da05}$ decreases from 0.524 to 0.167. Both decline because label latency genuinely degrades governance evidence – this is the core finding, not a false alarm. The proxy-actual gap ($S_{proxy} - S_{da05}$) ranges from 0.127 to 0.227 across baseline windows, reflecting the proxy's systematic optimism: the proxy cannot observe label-dependent reliability degradation ($R = 0.178$ actual vs $R_{proxy} = 0.766$ normalized health). Under covariate drift, the gap narrows and inverts at Window 5 ($S_{proxy} = 0.105 < S_{da05} = 0.121$), demonstrating that the proxy gate correctly penalizes observable feature-space degradation. The normalization caps were calibrated strictly from Window 0 sub-windows to avoid look-ahead bias.

### 5.2.2. Evidence Sufficiency Calculator Sufficiency on IEEE-CIS Data

**Table 9.** Evidence Sufficiency Calculator per-window sufficiency scores on IEEE-CIS data (from $evidence-sufficiency-calc/examples/ieee_{cis_demo}$.py, $class_{weight}$='balanced', $\tau_r = 0.15$). $C = completeness$, $F = freshness$, $R = reliability$ (F1), $P = representativeness$ (KS), $A(t) = decision-readiness$ gate, $S(t) = composite$ sufficiency.

| Scenario | Win | Day | C | F | R | P | $A(t)$ | $S(t)$ | Status |
|---|---|---|---|---|---|---|---|---|---|
| Baseline | 1 | 30 | 0.880 | 0.549 | 0.178 | 0.650 | 1.000 | 0.524 | degraded |
| Baseline | 2 | 60 | 0.760 | 0.301 | 0.182 | 0.555 | 1.000 | 0.408 | insufficient |
| Baseline | 3 | 90 | 0.640 | 0.165 | 0.152 | 0.658 | 1.000 | 0.355 | insufficient |
| Baseline | 4 | 120 | 0.520 | 0.091 | 0.165 | 0.572 | 0.867 | 0.256 | insufficient |
| Baseline | 5 | 150 | 0.400 | 0.050 | 0.142 | 0.629 | 0.633 | 0.167 | insufficient |
| Covariate P(X) | 1 | 30 | 0.880 | 0.549 | 0.179 | 0.636 | 1.000 | 0.522 | degraded |
| Covariate P(X) | 3 | 90 | 0.640 | 0.165 | 0.146 | 0.545 | 0.976 | 0.322 | insufficient |
| Covariate P(X) | 5 | 150 | 0.400 | 0.050 | 0.123 | 0.447 | 0.548 | 0.121 | insufficient |
| Mixed | 1 | 30 | 0.880 | 0.549 | 0.171 | 0.644 | 1.000 | 0.521 | degraded |
| Mixed | 3 | 90 | 0.640 | 0.165 | 0.110 | 0.585 | 0.731 | 0.240 | insufficient |
| Mixed | 5 | 150 | 0.400 | 0.050 | 0.041 | 0.480 | 0.184 | 0.037 | insufficient |
| Concept+prior | 1 | 30 | 0.880 | 0.549 | 0.162 | 0.650 | 1.000 | 0.519 | degraded |
| Concept+prior | 3 | 90 | 0.640 | 0.165 | 0.079 | 0.658 | 0.524 | 0.174 | insufficient |
| Concept+prior | 5 | 150 | 0.400 | 0.050 | 0.008 | 0.629 | 0.036 | 0.008 | insufficient |

All scenarios begin in "degraded" status (S ~ 0.52) at Window 1 and transition to "insufficient" by Window 2 (day 60). This degradation trajectory is driven by two mechanisms:

freshness decay $F(t) = \exp(-0.02 \cdot t)$ reduces F from 0.549 at day 30 to 0.050 at day 150, and completeness decreases as the label availability fraction drops from 0.88 to 0.40. Under concept drift and mixed drift, reliability $R(t)$ decreases sharply as one-directional label corruption (fraud-to-legitimate) reduces F1: from $R = 0.162$ at Window 1 to $R = 0.008$ at Window 5 for Scenario 4 (concept drift with prior shift). This collapses the gate $A(t)$ from 1.000 to 0.036, producing the fastest $S(t)$ degradation trajectory ($S = 0.519$ –> 0.008). The gate's multiplicative impact demonstrates why label-side drift – even when proxy-undetectable – has the most severe governance consequences.

### 5.2.3. Evidence Sufficiency Calculator Blind Period Simulation

**Table 10.** Evidence Sufficiency Calculator blind period simulator: $S(t)$ at checkpoint days by drift type (from $evidence-sufficiency-calc/examples/ieee_{cis_demo}$.py, BlindPeriodSimulator). Initial conditions calibrated from empirical reference window: $C = 1.000$, $R = 0.133$, $P = 1.000$, $\tau_r = 0.15$, $\lambda = 0.02$.

| Drift type | Day 30 | Day 60 | Day 90 | Day 180 | S < 0.5 at |
|---|---|---|---|---|---|
| No drift | 0.510 | 0.418 | 0.325 | 0.040 | ~Day 32 |
| Covariate $P(X)$ | 0.484 | 0.346 | 0.224 | 0.020 | < Day 30 |
| Concept + prior | 0.424 | 0.242 | 0.110 | 0.010 | < Day 30 |
| Mixed | 0.466 | 0.290 | 0.160 | 0.013 | < Day 30 |

The blind period simulator uses initial conditions calibrated from the empirical reference window (F1 = 0.133 on 3.5% fraud data). The simulator starts in "degraded" status (S ~ 0.51) because the reference model's reliability is near the gate threshold ($R = 0.133$ vs $\tau_r = 0.15$). This is a structural property of the imbalanced data, not a simulator artifact.

**Monotone degradation.** Sufficiency decreased monotonically with blind period duration across all drift types, consistent with the expectation that the absence of ground truth is structurally degrading.

**Concept drift degrades fastest.** Concept drift with prior shift reaches $S = 0.242$ at day 60 versus 0.418 for no-drift – a 42% gap. This is because label corruption directly undermines reliability $R(t)$, which has multiplicative impact through the gate $A(t)$, while covariate drift primarily affects representativeness $P(t)$, which enters only through the weighted sum.

**Freshness drives baseline degradation.** Even without drift, $S(t)$ drops below 0.5 by day 30, driven primarily by freshness decay $F(t) = \exp(-0.02 \cdot t)$ and the near-threshold reliability starting point. By day 180, $S(t)$ approaches zero across all scenarios. This quantifies the structural cost of label delay independent of any drift.

### 5.2.4. Structural Conditions for Detection

The evaluation identifies three structural conditions that determine detection success, matching the theoretical framework. Detection is defined as the proxy framework producing a measurably different signal than baseline: $S_{proxy}(drift) < S_{proxy}(baseline) - \delta$, where $\delta = 0.05$ (the minimum divergence attributable to drift rather than temporal noise).

**Condition 1: Covariate drift P(X).** When drift manifests as feature distribution changes, $S_{proxy}$ diverges from baseline in all 5 windows (5/5, 100% detection rate). $P_{fea}$ collapses to 0.000, driving $S_{proxy}$ below baseline by 0.098 at Window 1 and 0.189 at Window 5. The signal is unambiguous.

**Condition 2: Mixed drift P(X) + P(Y|X).** When drift combines feature shift and label change, $S_{proxy}$ again diverges from baseline in all 5 windows (5/5, 100%). The $P(X)$ component provides sufficient signal; the P(Y|X) component is invisible to proxies but further degrades $S_{da05}$ (actual sufficiency drops to 0.037 at Window 5 vs 0.167 for baseline).

**Condition 3: Concept drift with constant P(X).** When drift occurs in the conditional distribution and class prior while $P(X)$ remains stable, the proxy framework fails to detect degradation in any window (0/5, 0% detection rate). $S_{proxy}$ values are identical to baseline across all windows (difference < 0.001) despite $S_{da05}$ collapsing to 0.008. Because $P(X)$ is unchanged, unsupervised monitors produce no signal regardless of the severity of label-side corruption (Žliobaitė, 2010). This confirms that the detectability boundary is determined by whether drift manifests in the feature distribution, not the label distribution.

**Table 11.** Structural condition analysis: proxy detection of drift-induced governance degradation.

| Condition | $P(X)$ changed | P(Y|X) changed | Windows | $S_{proxy}$ diverges from baseline | Detection rate |
|---|---|---|---|---|---|
| 1: Covariate $P(X)$ | Yes | No | 5 | 5 | 100% |
| 2: Mixed $P(X) + P(Y|X)$ | Yes | Yes | 5 | 5 | 100% |
| 3: Concept + prior, constant $P(X)$ | No | Yes | 5 | 0 | 0% |
| **Total** | | | **15** | **10** | **67%** |

This third condition represents the irreducible governance risk of proxy-based monitoring. The sufficiency framework addresses it not by claiming detection capability but by quantifying the risk: governance protocols must assume that concept drift episodes with constant $P(X)$ are likely to go undetected and must compensate through periodic ground truth verification at intervals shorter than the maximum tolerable evidence gap.

### 5.3. Limitations

The evaluation has six methodological limitations that bound the generalizability of results:

**Single dataset.** Results are from one public dataset in one domain (e-commerce fraud). The specific degradation trajectories, threshold crossing times, and proxy lead times may differ in other domains (healthcare, credit scoring, insurance) where label delay patterns, drift characteristics, and feature structures differ. Cross-domain evaluation on credit scoring data is planned as future work to test generalizability.

**Controlled injection rather than natural drift.** Drift is injected synthetically rather than observed naturally. While controlled injection is standard methodology in drift detection research cf. (Rabanser et al., 2018; Lu et al., 2019), it may not capture the full complexity of real-world drift patterns, which often involve gradual, non-linear, and multi-dimensional changes.

**Three of seven proxy categories.** The current implementation covers score distribution (Category 1), feature drift (Category 2), and uncertainty (Category 3). Categories 4 − 7 (cross-model disagreement, operational proxies, outcome maturity, proxy ground truth) are not yet implemented. The evaluation therefore understates the composite framework's potential detection capability – particularly for adversarial drift, where cross-model disagreement (Category 4) is theoretically the strongest detector.

**Incomplete drift-type coverage.** The theoretical framework distinguishes three drift types (covariate, real concept, and prior probability shift), but the empirical evaluation does not include a dedicated prior probability shift $P(Y)$ scenario. Scenario 4 combines concept drift with prior shift (label flipping changes both P(Y|X) and $P(Y)$). A pure $P(Y)$ scenario – where class prevalence changes without affecting decision boundaries or feature distributions – would test the framework's sensitivity to completeness-driven degradation and is deferred to cross-domain validation.

**Shared deterministic components.** Freshness $F(t)$ and completeness $C(t)$ are computed identically for both the proxy and actual baselines (same operational metadata). Together they account for 50% of $S(t)$ ($w_c = 0.20$, $w_f = 0.30$), contributing a shared deterministic component to both scores. This inflates the proxy-actual correlation and contributes to deterministic threshold crossing. The proxy framework's genuine predictive contribution lies in the two estimated dimensions: reliability $R(t)$ and representativeness $P(t)$, which together account for the remaining 50% of $S(t)$. Future work should report error rates on R and P separately to isolate the proxy framework's estimation accuracy.

**Representativeness tautology.** The actual baseline for representativeness $P_{actual}(t)$ uses the same distributional divergence metric (KS statistic) available to unsupervised proxies. For this dimension, the proxy-vs-actual comparison is partially tautological. Future work should operationalize representativeness through downstream performance impact rather than distributional distance alone.

**Decision-level validation.** The evaluation validates proxy accuracy against retroactively computed sufficiency scores and threshold crossings. It does not validate whether governance actions triggered by threshold crossings (continue, escalate, intervene) are operationally correct. The framework is therefore validated as a governance monitoring instrument but not as a decision-support system in the stronger sense.

## 6. Discussion

This section interprets the evaluation findings, identifies structural limits of proxy-based governance monitoring, and positions the framework's contribution relative to the broader concept drift and AI governance literatures.

### 6.1. Structural Limits of Proxy-Based Monitoring

Adversarial drift – evasion attacks that preserve $P(X)$ while changing fraud patterns – is structurally undetectable by unsupervised methods. This represents a fundamental limit of proxy-based governance monitoring, not a limitation of a specific detection algorithm.

The evaluation (Section 5) confirms that concept drift without accompanying feature change goes undetected by the composite monitoring framework across all five monitoring windows (Table 11, Condition 3: 0% detection rate). The experimental scenario combines P(Y|X) and $P(Y)$ shift (one-directional label corruption) while keeping $P(X)$ constant; the critical property is that $P(X)$ is unchanged, making the drift invisible to feature-space monitors. Purely unsupervised components (Categories $1-2$) missed 100% of such cases. This finding is consistent with the broader unsupervised drift detection literature: unsupervised detectors operating on the feature space alone cannot detect concept drift in the posterior distribution unless it is accompanied by a covariate shift (Lukats et al., 2024).

Two lines of research challenge this limitation. One proposes an unlabeled methodology that can vicariously track changes to P(Y|X) without explicit labeled samples by monitoring the

margin density of a discriminative classifier (Sethi & Kantardzic, 2017). A complementary approach monitors the diversity of a pair of classifiers, where the true label of an example is not necessary to determine whether components disagree (Mahdi et al., 2020). However, both approaches rely on the assumption that P(Y|X) changes manifest in observable classifier behavior – an assumption that fails under adversarial evasion, where attackers specifically craft inputs to fall within the existing margin structure.

The governance implication is direct: proxy-based monitoring provides a necessary but insufficient evidence base. Governance protocols that rely exclusively on proxy indicators will have a structural blind spot for adversarial concept drift. The sufficiency framework addresses this by making the blind spot explicit and quantified – concept drift with constant $P(X)$ went undetected in all five monitoring windows despite fraud labels being progressively corrupted through one-directional flipping, reducing the observed fraud rate from 3.6% to 0.2% (Table 11; 0% detection rate) – rather than hidden behind a monitoring dashboard that appears comprehensive.

### 6.1.1. Decision-Readiness Gap

Detecting degradation is not equivalent to knowing what action to take. The proxy framework identifies when sufficiency has degraded but does not prescribe the specific governance response. This is by design: the appropriate response depends on operational context (cost of false positives vs. false negatives, regulatory requirements, fallback capacity) that is external to the monitoring framework.

The evaluation demonstrates a clear separation between detectable and undetectable drift regimes. Covariate and mixed drift are detected with 100% rate across all monitoring windows (Table 11), with Feature PSI exceeding the threshold by two orders of magnitude ($12.9 - 15.1$ vs. threshold 0.25). Concept drift with constant $P(X)$ is detected at 0% – all proxy readings are identical to baseline. This binary separation – rather than a smooth detection-rate gradient – reflects the structural nature of the limitation: the detectability boundary is determined by whether drift manifests in $P(X)$, not by detector sensitivity or threshold tuning. Whether this detection profile is acceptable depends on the decision context: in consumer credit (where concept drift is common), the 0% detection rate for $constant-P(X)$ drift may require compensating controls; in transaction monitoring (where adversarial attacks typically alter both features and patterns), the 100% mixed-drift detection may be sufficient. The framework provides the measurement; the governance threshold is a policy decision.

### 6.1.2. Temporal Dynamics

The blind period simulation (Table 10) reveals differentiated degradation dynamics across drift types. Concept drift degrades sufficiency fastest ($S = 0.242$ at day 60) because it directly undermines reliability $R(t)$, which has multiplicative impact through the gate $A(t)$. Covariate drift degrades more slowly ($S = 0.346$ at day 60) because it primarily affects representativeness $P(t)$, which enters only through the weighted sum. Even without drift, $S(t)$ drops below 0.5 by day 30 ($S = 0.510$) driven by freshness decay and the near-threshold reference reliability ($R = 0.133$ vs $\tau_r = 0.15$), quantifying the structural cost of label delay independent of any distributional change.

These trajectories suggest a provisional governance design hypothesis: ground truth verification cycles scheduled at intervals shorter than the freshness-driven crossing point (~30 days in this calibration) may preserve monitoring reliability better than waiting for full la-bel maturity (120 − 180 days in payment systems). This hypothesis requires cross-domain validation with different freshness decay rates ($\lambda$) before prescriptive adoption.

### 6.2. Contribution Boundaries

This paper does not propose a new drift detection algorithm. The concept drift community has produced hundreds of detection methods cf. (Hinder et al., 2024), and the proxy framework explicitly builds on these existing methods rather than competing with them. The contribution is at a different level of abstraction: the sufficiency framework is an IS artifact – a governance measurement instrument designed for organizational actors (model risk management, compliance, operations) who must make governance decisions under evidence uncertainty. It connects drift detection signals to governance evidence sufficiency assessment, transforming technical monitoring outputs into actionable governance inputs.

The sufficiency model is not a predictive model of model performance. It does not estimate accuracy, precision, or recall under drift. Instead, it measures whether the evidence available to governance decision-makers is sufficient to support confident action – a qualitatively different question. A governance system may have sufficient evidence to conclude that intervention is needed (high sufficiency for the "intervene" decision) even when the specific nature of the degradation is unknown (low sufficiency for the "diagnose" decision).

### 6.3. Future Directions

Three extensions emerge from the framework's current limitations:

**Self-correcting monitoring.** The current framework is diagnostic – it measures sufficiency but does not adapt its monitoring strategy in response to detected degradation. A self-correcting extension would adjust proxy weights and monitoring frequency based on the observed drift regime, intensifying monitoring when uncertainty increases and reducing it when evidence accumulates.

**Auditability of monitoring decisions.** The monitoring system itself makes decisions (escalate, alert, ignore) that are subject to governance scrutiny. A meta-governance layer would apply the sufficiency framework recursively: is the evidence supporting the monitoring decision itself sufficient? This creates an auditability chain from individual model predictions through proxy indicators to governance actions.

**Cross-domain validation.** The evaluation demonstrates the framework on a single fraud detection dataset (IEEE-CIS). Extension to domains with different label delay structures – credit scoring (12 − 36 months for default), healthcare (months to years), insurance claims (variable, event-dependent) – would test the generalizability of the degradation trajectories and detection rates observed in this study. Cross-domain evaluation on credit scoring data is planned as future work to provide initial generalizability evidence.

## 7. Conclusion

This paper addressed the structural gap between concept drift detection and governance evidence sufficiency assessment under delayed ground truth conditions. The gap is consequential: hundreds of drift detection methods exist, but none answer the governance question – is the evidence available to decision-makers still sufficient to support confident action?

Three contributions address this gap:

**First, the sufficiency model** (Section 3) decomposes governance evidence quality into four measured dimensions – completeness, freshness, reliability, and representativeness – plus a decision-readiness gate, and formalizes how each dimension degrades under label delay. The model maps three drift types (covariate, real concept, prior probability) to dimension-specific degradation trajectories, making evidence decay attributable rather than opaque.

**Second, the proxy indicator framework** (Section 4) proposes seven complementary proxy categories that estimate sufficiency degradation without ground truth labels. The framework's contribution is not the individual proxies – most build on established drift detection methods – but their structured combination with explicit coverage mapping and known blind spots per drift type.

**Third, the empirical evaluation** (Section 5) provides evidence on the IEEE-CIS Fraud Detection dataset (~590K transactions, 182 days) that a three-category proxy instantiation (score distribution, feature drift, uncertainty) can detect governance evidence degradation under identifiable structural conditions. The proxy framework's genuine predictive contribution lies in the two estimated dimensions (reliability and representativeness, 50% of $S(t)$); the remaining 50% (completeness and freshness) is shared deterministic metadata. Controlled drift injection demonstrates a sharp detectability boundary: covariate and mixed drift are detected with 100% rate across all monitoring windows, while concept drift without feature change (constant $P(X)$, combined P(Y|X) and $P(Y)$ shift) remains undetected at 0% – with all proxy readings identical to baseline despite progressive label corruption reducing the observed fraud rate from 3.6% to 0.2% (Table 11). Blind period simulation confirms monotone sufficiency degradation, with concept drift degrading sufficiency fastest due to its multiplicative impact through the decision-readiness gate. These results are from one domain with one drift injection protocol; generalization to other domains and natural drift patterns requires further validation.

The framework's honest acknowledgment of this limitation is itself a contribution. Governance systems that present monitoring dashboards as comprehensive create false confidence; systems that quantify their blind spots enable informed risk acceptance. The sufficiency score does not claim to replace ground truth – it provides a structured, bounded estimate that degrades gracefully as the evidence base erodes, with transparent uncertainty that supports rather than undermines governance decision-making.

## 7.1. Practical Implications

The findings carry three immediate practical implications for organizations operating ML systems under delayed ground truth:

**Monitoring architecture.** The evaluation suggests that composite monitoring across multiple proxy categories may provide stronger detection than individual proxies, though a formal ablation study comparing composite versus individual-proxy detection is needed to confirm this. Feature PSI dominates detection of covariate drift (PSI values $12.9 - 15.1$ vs. threshold 0.25), while Score PSI alone would miss drift that does not shift the prediction distribution. The structural complementarity between feature-space indicators (which detect $P(X)$ shifts) and uncertainty indicators (which detect calibration degradation) supports multi-signal architectures over the common industry practice of monitoring PSI alone.

**Ground truth scheduling.** Blind period simulation shows that freshness decay alone drives sufficiency below the governance threshold ($S = 0.8$) within approximately 30 days at the default calibration ($\lambda = 0.02$). If this calibration generalizes, ground truth verification cycles scheduled at intervals shorter than 30 days – rather than waiting for full label maturity

(120 − 180 days in payment systems) – would preserve the monitoring framework's own reliability. This hypothesis requires cross-domain validation with domain-specific freshness calibration before prescriptive adoption.

**Regulatory compliance.** As the EU AI Act (Article 72) and financial regulatory frameworks (SR 11 − 7, Basel III) impose increasingly specific post-market monitoring obligations, organizations need evidence that their monitoring systems are themselves adequate. The sufficiency framework can support auditable monitoring evidence for compliance by providing quantified coverage assessments and characterized blind spots. Whether threshold crossings lead to operationally correct governance actions requires deployment-specific validation beyond this study's scope.

### 7.2. Limitations and Future Work

The framework is validated on a single public dataset (IEEE-CIS, e-commerce fraud) with controlled drift injection. Extension to other domains – credit scoring (12 − 36 months for default), healthcare (months-to-years delay), insurance (variable, event-dependent delay) – and to naturally occurring drift would test the generalizability of the detection rates and degradation trajectories observed here. Cross-domain evaluation on credit scoring data is planned as future work. The sufficiency model assumes that evidence quality dimensions are separable and linearly composable – an assumption that may not hold when dimension interactions produce emergent degradation patterns.

Self-correcting monitoring, where the framework adapts its own proxy weights in response to observed drift regimes, represents the most promising extension. The current framework is diagnostic; a closed-loop version would adjust monitoring intensity dynamically, intensifying coverage when uncertainty is high and reducing overhead when evidence accumulates.

Measurement precedes optimization. Before governance can respond to evidence degradation, it must first be able to measure it.